# ReS$_2$/h-BN/Graphene Heterostructure Based Multifunctional Devices: Tunnelling Diodes, FETs, Logic Gates & Memory

Bablu Mukherjee[1*], Ryoma Hayakawa[1], Kenji Watanabe[2], Takashi Taniguchi[1], Shu Nakaharai[1], Yutaka Wakayama[1*]

[1]International Center for Materials Nanoarchitectonics (MANA), National Institute for Materials Science (NIMS), 1-1 Namiki, Tsukuba, Ibaraki 305-0044, Japan.

[2]Research Center for Functional Materials, National Institute for Materials Science (NIMS), 1-1 Namiki, Tsukuba, Ibaraki 305-0044, Japan.

* Corresponding authors. Email: MUKHERJEE.Bablu@nims.go.jp (Bablu Mukherjee); WAKAYAMA.Yutaka@nims.go.jp (Yutaka Wakayama)

## Abstract

We investigate a two-dimensional (2D) heterostructure consisting of few-layer direct bandgap ReS$_2$, a thin h-BN layer and a monolayer graphene for application to various electronic devices. Metal-insulator-semiconductor (MIS)-type devices with two-dimensional (2D) van-der-Waals (vdW) heterostructures have recently been studied as important components to realize various multifunctional device applications in analogue and digital electronics. The tunnel diodes of ReS$_2$/h-BN/graphene exhibit light tuneable rectifying behaviours with low ideality factors and nearly temperature independent electrical characteristics. The devices behave like conventional MIS-type tunnel diodes for logic gate applications. Furthermore, similar vertical heterostructures are shown to operate in field effect transistors with a low threshold voltage and a memory device with a large memory gate for future multifunctional device applications.

# I. Introduction

An assembly of several two-dimensional (2D) crystals via a weak van-der-Waals (vdW) interaction in one vertical stack exhibits various exciting physical phenomena.[1-6] The unique electronic transport properties in the 2D heterostructures via band alignment modulation, is the key to utilize them in multifunctional high performance electronic devices.[1,2] The use of a vdW heterostructure with various atomically flat 2D materials makes it possible to realize various electronic primary components such as diodes, transistors, capacitors and memory devices, which would make it possible to integrate multifunctional devices on a single chip. 2D-heterostructure-based metal-insulator-semiconductor (MIS)-type devices have various potential applications in analogue and digital electronics.[3,4] For instance, insulating atomically flat hexagonal boron nitride (h-BN) can function as a 2D tunnelling layer between a conductor and a semiconducting channel thus allowing perfectly planar charge injection across the atomically flat interface without various charge trapping sites derived from dangling bonds.[5,6] Such quantum tunnelling can make it possible to realize vertical MIS-type two-terminal diodes with the superior and temperature independent functionality, which is an advantage for integrated circuit technology.[7-10] Thus the heterostructure of vdW materials allows to create vertical diodes for logic gate operation in future digital electronics. The opportunity for increasing the integration density is expected to be greater in vertically stacked MIS-type-tunnel-diode-based logic gates circuits. Compared with conventional Si 3D semiconductor-based electronic devices, all-2D layered materials and their heterostructures are expected to be superior for application to quantum tunnel diodes with high performance and less temperature dependence.

Field-effect transistors (FETs) and non-volatile memory (NVM) devices have been studied experimentally and theoretically because large-scale CMOS compatible three-terminal FETs are convenient for application to integrated circuit technology. A highly insulating h-BN layer can be used as a 2D dielectric layer for $ReS_2$ FET operation with atomically flat graphene as a local gate, where the strong electrostatic modulation of $ReS_2$ channel conduction can be achieved. This can efficiently reduce the operating bias for $ReS_2$ FETs with a reasonable ON/OFF ratio for low power operation. In another application to a memory device, a graphene layer isolated from the $ReS_2$ channel with insulating h-BN can be used as a floating gate for charge storage. However, many challenges and problems still remain to be overcome before we can realize real device applications, and these include temperature dependence, high operating voltage, low operation speed, poor reliability, low integration density, low optical absorption and poor sensing bandwidth. Recent studies with direct bandgap multilayer $ReS_2$ (~ 1.4 – 1.5 eV) have reported various electronics and optoelectronics applications.[11-15] Different MIS-type device structures have been explored using 2d material based heterostructures[3, 16-20], as has their device application[21] and logic gate operation[9, 22-24]. However, little work has been reported on $ReS_2$ based heterostructure in relation to MIS-type tunnel diodes with a view to using these diodes for logic gate operation and the heterostructure for multifunctional device applications. By contrast, for transistor operation, the ON/OFF current ratio is always compromised due to the application of a low gate field, and for memory operation it is always critical to obtain a large memory window for future multilevel memory operation in ultrathin nanoelectronics.

The main purpose of this study is to build multifunctional devices using 2D-layered vdW materials as basic building blocks for diodes, for field effect transistors (FETs) operating at a low bias, and for logic gate applications and non-volatile memory (NVM) applications. Different layered material components were included in the vdW stack via a multistep transfer process to realize multifunctional device applications. Recently, we have developed a laser-assisted non-

Article type: Full Paper

volatile memory device for multi-level storage applications using a heterostructure consisting of various 2D materials.[25] A similar heterostructure device can be used for logic gate device applications, which further expands the possibilities to multi-functional device applications. Quantum tunnelling phenomena can be incorporated in a designed heterostructure in order to obtain almost temperature independent operation in the fabricated diodes. $ReS_2$/h-BN/graphene vertical MIS-like heterostructures as quantum tunnelling diodes (QTDs) are studied in detail by characterizing the gate electric field dependence, temperature dependence and incident light excitation. We then realize logic gates operation by using a vertical MIS-type tunnel diode. The vertical geometry can lead to tunnel devices with a high rectification ratio and an atomically flat 2D surface as high tunnelling area for a large turn-ON current and a low turn-ON voltage. The fabricated diodes were realized by direct tunnelling (DT) at a lower field strength and Fowler-Nordheim (FN) tunnelling by the application of a high-strength electric field. These tunnel diodes can be used to design various logic gates. We obtained a large rectification, low ideality factor, temperature independent tunnel diodes and their logic gate operation. Furthermore, similar heterostructure geometry can be used for other active electronic devices such as low operating bias FETs transistors and memory devices by arranging different electrical contacts, which are described in the second part of this work. The key parameter of achieving multifunctionality of the device is the fine tunning of the h-BN thickness. The reduction of h-BN layer thickness enables low bias operating FETs, high gate bias modulated NVM operation and vertical MIS-type tunnel diodes. We found that h-BN thickness of 3-5 nm is suitable for the both tunneling, namely DT and FN-tunneling, leading to applications like vertical tunnel MIS-type diodes. Whereas, h-BN thickness of 6-10 nm is applicable for FN-tunneling, where low bias operating FETs and high gate bias modulated NVM devices can be fabricated. This study provides an overview of the use of all-

Article type: Full Paper

2D atomically-flat layered materials to realize various electronic components including diodes, gate-tunable diodes, low bias operating FETs, logic gate devices and memory devices for multifunctional electronic device applications. These devices are candidates for use in the future smart technology Internet-of-Things (IoT) as building blocks in next generation VLSI architectures, which can potentially enable all-opto-electrical logic processing and quantum information processing.

## II. Results and Discussion

## II. a. MIS-type Diode Performance & Logic Gate Operation:

A MIS-type diode structure for logic gate application is a vdW heterostructure consisting of multilayer $ReS_2$ supported by high-quality thin h-BN as a tunnelling layer, where we placed monolayer graphene (Gr) beneath the h-BN as a conductive layer. The 2D nanosheets were directly exfoliated on top of a polydimethylsiloxane (PDMS) stamp and transferred to the top of another nanosheet on the $SiO_2$/Si substrate. The standard dry transfer method was used to transfer various large-area high-quality 2D flakes onto an arbitrary substrate. The device fabrication is described in detail in the experimental section. In a vdW heterostructure with $ReS_2$ as a semiconducting layer will have to reduce scattering from the charged impurities of the substrate and interfacial impurities resulting from the presence of the atomically flat vdW material h-BN. **Figure 1a-c**, respectively, show an optical image of the different components used to fabricate the heterostructure, and AFM images of the insulating layer h-BN and the top $ReS_2$ layer obtained with a line scan.

Article type: Full Paper

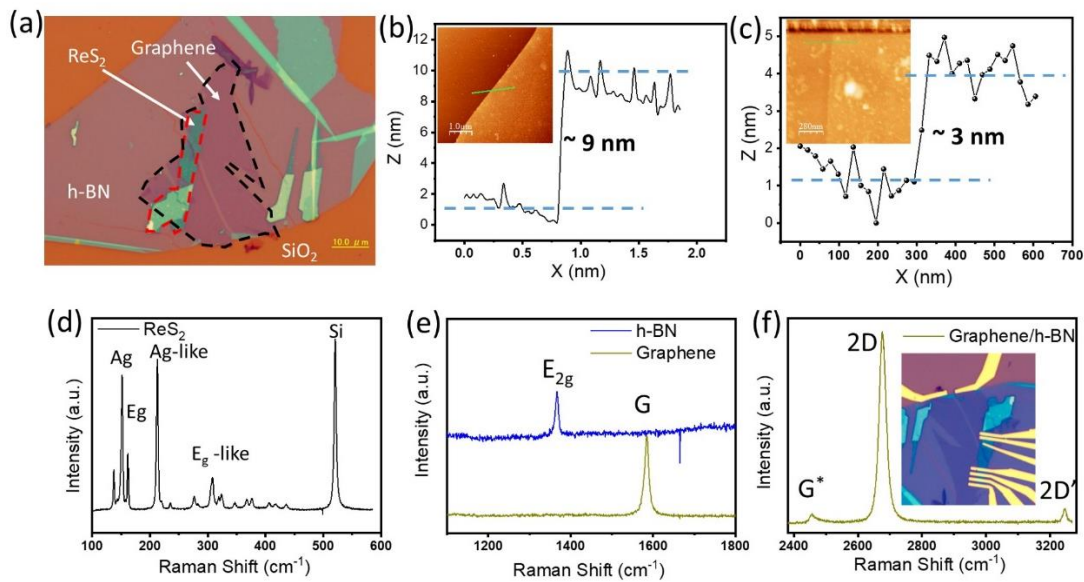

**Figure 1.** (a) Optical image of a fabricated multilayer ReS$_2$/h-BN/graphene heterostructure before the addition of electrical contacts. (b, c) Height profile corresponding to the solid green line in the inset surface topography AFM image for h-BN and ReS$_2$ layers, respectively. (d) Raman spectra taken from the top surface of the ReS$_2$ flake in the ReS$_2$/h-BN/graphene heterostructure. (e, f) The Raman spectrum was collected from the region corresponding to the h-BN and graphene/SiO$_2$, respectively. The inset shows an optical image of the device after the addition of electrical contacts.

We confirmed the crystalline quality of the transferred ReS$_2$ layer by Raman spectroscopy (**Figure 1d**), which indicated the high crystallinity and high chemical purity of the ReS$_2$ flake material. The ReS$_2$ layer was a few atomic layers (~3 nm) thick, which was further verified with an AFM scan and with a line profile monitoring the thickness. We observed typical E$_{2g}$ (in-plane vibration mode) and A$_{1g}$ (out-of-plane vibration mode) peaks at 162 and 212 cm$^{-1}$, respectively, along with other labelled peaks. The Raman spectrum of graphene (**Figure 1e,f**) exhibits a highly crystalline monolayer signature as it has a 2D/G peak integral intensity ratio greater than 2 and a narrow (and symmetric) 2D peak width (FWHM) of 22.1 ± 0.5 cm$^{-1}$ at 2680.4 cm$^{-1}$. Other Raman peak signatures were present, namely that of an E$_{2g}$ in-plane phonon caused by the CC stretching

Article type: Full Paper

vibration mode and of a D peak at 1350 cm$^{-1}$, which is related to defects. The h-BN signature was observed in the $E_{2g}$ peak, which is the in-plane phonon vibration mode (**Figure 1e**) at 1366.4 cm$^{-1}$, where the thickness was ~ 9 nm as monitored from the AFM topography image and the line profile (**Figure 1 e,f**). Choosing the right h-BN thickness is very important in terms of promoting direct tunnelling (DT) and/or Fowler - Nordheim (FN) tunnelling and we employed h-BN layers with a thickness of 5-9 nm for both electron and hole tunnelling. Varying the thickness of h-BN can make it possible to realize various different applications including MIS-type diodes, FETs with a low threshold voltage and memory applications, which are discussed in subsequent sections. It is worth noting that various defect levels and/or trap levels have also been reported in h-BN layers and ReS$_2$, which might also affect the tunnelling process.[11, 15, 26-30]

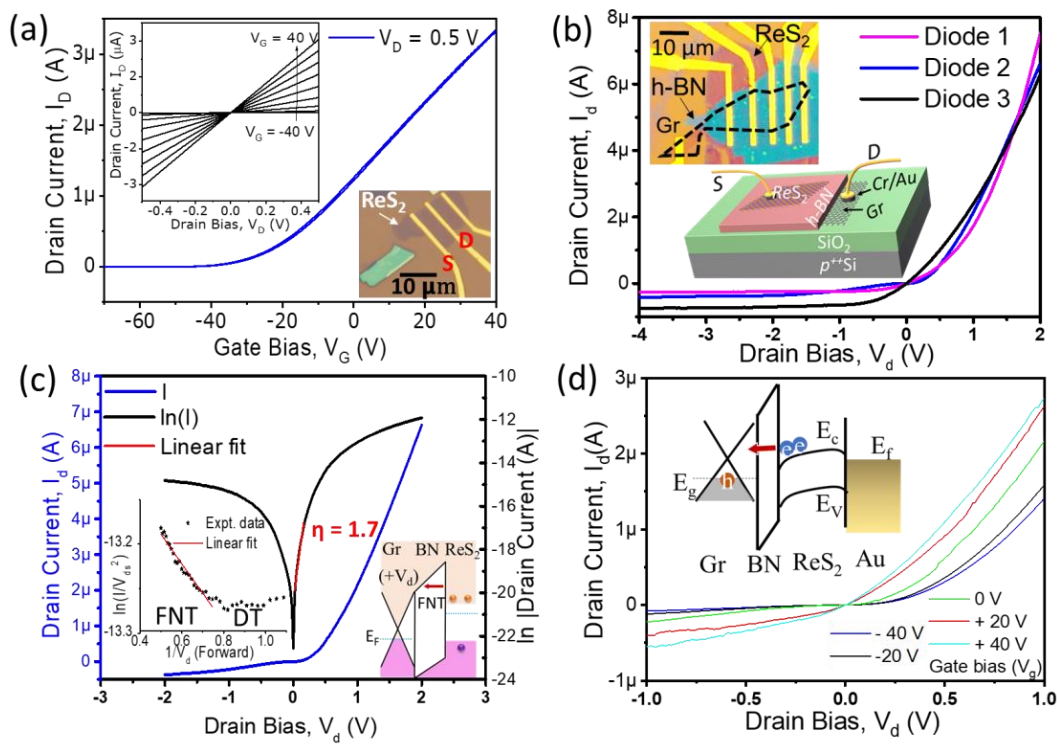

**Figure 2.** (a) A typical transfer curve ($I_d$ versus $V_g$) of an ReS$_2$ FET device, showing n-type characteristics. The insets show the output curve with the SiO$_2$ back gate $V_g$ varying in 10 V steps

and an optical image of the ReS$_2$ FET. Device performance of MIS-type diode: (b) Output curve measured across the heterostructures for three different diodes. The insets show an optical image of the tested device and a schematic illustration of the MIS-type diode. (c) A linear and log-scale plot with diode equation fitting that shows an ideality factor of 1.7. The inset shows the FNT-curve fitting in the high field region and a schematic band diagram obtained during FN-tunnelling. (d) Output characteristics curves: drain current (I$_d$) versus V$_d$ for different control gate biases V$_g$. The inset shows the energy band diagram of an ReS$_2$/h-BN/graphene heterostructure with a positive drain voltage at a graphene terminal.

First, we tested $p^{++}$Si back-gated few-layer ReS$_2$ channel-based FET devices (Inset **Figure 2a**) at room temperature to identify the majority carrier transport using the conventional solid gate dielectric. The representative transfer characteristics ($I_d$-$V_g$; sweep V$_g$) and output characteristics ($I_d$-$V_d$) for different V$_g$ values are represented in **Figure 2a**, and they suggest that ReS$_2$ has n-type conductance with natural electron doping. The linear output characteristics ($I_d$-$V_d$; inset **Figure 2a**) shows effective ohmic contacts with the Cr/Au electrodes. The electrical performance of the MIS-type diodes, namely the current versus voltage (*I-V*) curves, was measured across the heterostructure device consisting of a few-layer ReS$_2$/h-BN/graphene to demonstrate the formation of vertical MIS-type diodes with high rectification and a low turn-on voltage. We evaluated the electrical characteristics (*I-V*) of the fabricated diodes as shown in the optical image in the **Figure 2b** inset. An h-BN tunnel barrier thickness of ~5 nm was used for MIS-type diodes application. During the electrical probe measurements, the graphene was connected to the drain electrode and the n-type ReS$_2$ was connected to the source electrode. Here, the graphene layer was found to be initially hole doped from the absorption of oxygen and/or water molecules from the air.[25] Typical

Article type: Full Paper

*I-V* characteristics of fabricated diodes show that all the MIS-type diodes exhibited a highly rectifying feature (**Figure 2b**). Room temperature output characteristics with different bias voltages are shown in **Figure S1**, where electron accumulation and hole inversion in the ReS$_2$ layer are clearly observed. A small ideality factor ($\eta$) of 1.7 (**Figure 2c**) was extracted from the forward-bias characteristics after fitting with the standard n–p diode forward-bias equation:

$$I_d = I_s \left( e^{qV_d/\eta k_B T} - 1 \right) \tag{1}$$

where I$_s$, q, k$_B$ and T represent the reverse-bias saturation current, electron charge, Boltzmann constant and absolute temperature of the junction, respectively. This suggests that high quality interfaces were formed between those ReS$_2$, h-BN and graphene layers with minimal interface recombination/generation of carriers. We also found that the *I-V* characteristics of the diodes are less likely to be temperature dependent (**Figure S2**). Notably, we can fit the Fowler-Nordheim tunnelling (FNT) relation in the high electric field region of the tunnelling current-voltage curve. The FNT equation is expressed as follows:

$$\ln \frac{I(V)}{V_{ds}^2} = ln \frac{A\,q^3 m}{8\pi h m^* \varphi_B d^2} - \frac{8\pi d \varphi_B^{\frac{3}{2}} \sqrt{2m^*}}{3hqV_{ds}} \tag{2}$$

where A, $\varphi_B$, q, m, m*, d and h are the effective contact area, barrier height, electron charge, free electron mas, effective electron mass, barrier width of the tunnelling layer of h-BN and Planck's constant, respectively. The FNT equation (**Equation 2**) fitted (Inset **Figure 2c**) under the condition of a forward bias tunnelling current in a high applied electric field is shown in the linear region of the ln(I(V)/V$^2$) vs. (1/V) plot.[17] The electron Schottky barrier height between graphene and h-BN ($\varphi_{\text{e-graphene}}$) of 1.2 eV, was extracted from the FNT equation fitted in the forward bias region, which is lower than the reported value of 2.7 eV.[17] This analysis shows the possibility of FNT-tunnelling based electronic conduction in the vertical MIS-type diode in a high applied field with the assistance of defects. Various defect levels and/or trap levels of h-BN could be incorporated[11, 15,]

Article type: Full Paper

[26-30] in the heterostructure device to promote increased tunnelling current in the vertical direction to further increase the ON current, which could be the reason for obtaining the low barrier height value. One can provide more direct evidence to distinguish DT and FNT by local excitation with a focused laser beam.[17] Furthermore, we modulate the diode performance by using a gate induced electric field and/or varying the incident laser intensity as follows. We applied a back-gate voltage (control gate bias $V_g$) to the degenerately doped silicon substrate ($p^{++}$Si) to tune the tunnel current *versus* voltage characteristics via electrostatic coupling and the systematic shifting of the Fermi level of the $ReS_2$ channel along with band alignment across the $ReS_2$/h-BN/graphene heterostructure. The representative output characteristics ($I_d$-$V_d$) of the FET structure with different $V_g$ values (-40 V to +40 V) showed a non-linear drain current, representing good rectifying behaviour, and gate-induced modulation of drain current was observed (**Figure 2d**). The flat band alignment of $ReS_2$/h-BN/graphene has type-II band alignment, and the positive forward bias band alignment of the structure is shown in the inset in **Figure 2d**. It is a scenario whereby under a forward bias condition, the majority of the carrier electrons move from the $ReS_2$ tunnel to the graphene to form a high drain current. Whereas with a reverse bias, the electrons tunnelling from graphene to $ReS_2$ are suppressed by depletion in $ReS_2$ at a positive bias resulting in the production of less tunnel current. Under dark condition, the tunneling current from graphene to $ReS_2$ is dominated due to high carrier density in graphene, whereas under light illumination the tunneling current is dominated from $ReS_2$ to graphene due to highly increase photo-generated carriers in $ReS_2$. When positive drain voltage is applied to the graphene electrode, the photo-generated carrier (i.e. majority carrier electron) from $ReS_2$ can tunnel efficiently to graphene, which can be seen from the band alignment (**Figure S3**), which produce high photocurrent in the MIS-type diode structure.

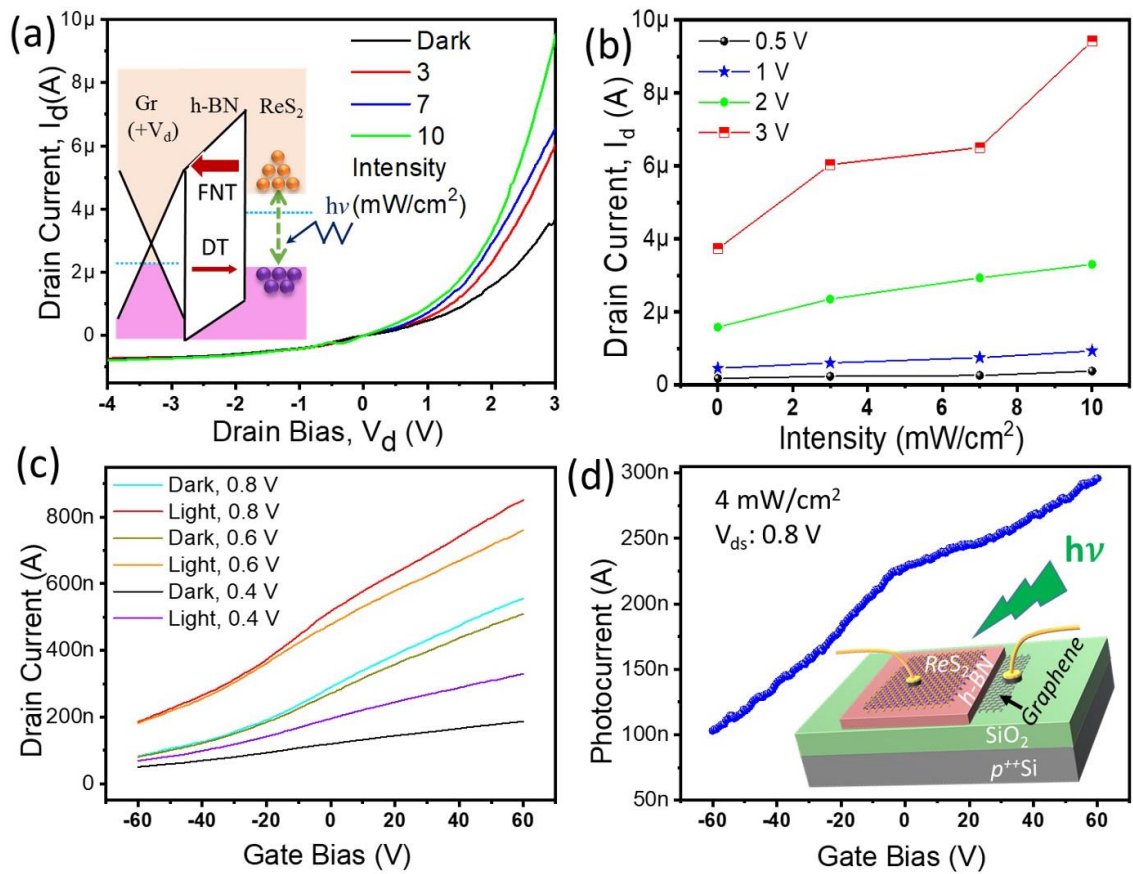

**Figure 3.** Optoelectrical characteristics of a device with graphene at the drain terminal: (a) Output characteristics curves: drain current ($I_d$) versus drain bias ($V_d$) under dark conditions and with 532 nm laser illumination of varying intensity. The inset shows an energy band diagram of the device with a positive voltage at the graphene terminal. The thickness of the red arrow indicates the amount of carrier tunneling. (b) Photocurrent *versus* laser intensity with different $V_d$ values. (c) $I_d$ *versus* $V_g$ with different $V_d$ values under dark conditions and with a fixed laser intensity of 4 mW/cm$^2$. (d) Photocurrent *versus* $V_g$ at a fixed drain voltage of 0.8 V at the graphene terminal. The inset shows a schematic of the heterostructure device under 532 nm light illumination.

We investigated the optoelectrical conduction process in the MIS-type diode structure with the possibility of light-assisted modulation in the rectification performance. Furthermore, the

Article type: Full Paper

optical excitation might result in a new functionality for tuning the device application in digital optoelectronics. Here we found that visible laser illumination (532 nm) produced a photocurrent, which was gate tuneable and the photocurrent generation mainly depended on the photoconductive gain in the semiconducting $ReS_2$, which is almost linear power law dependent in a low bias operation. The turn-on voltage of the diodes is modulated via different light intensities (**Figure 3a**). Rectification ratio under different light intensity is provided in **Table 1**(S.I.). The heterostructure forms a type-II band alignment (**inset Figure 3a**) with the top optically active $ReS_2$ layer. Photo-excited carrier generation at the $ReS_2$ semiconducting layer and carrier separation across the tunnel layer of h-BN to the conducting material graphene make further tuning the diode performance, which is more prominent in the FN-tunnelling region. Thus, in the positive high drain bias (forward bias) region we have observed a high photogenerated current (**Figure 3a**), which has an almost linear power dependence at a different bias (**Figure 3b**). On the other hand, possible defect and/or trap states located between the graphene Fermi level and the $ReS_2$ valence band enable hole transfer from graphene to $ReS_2$ via the h-BN layer. In one possible scenario the accumulated holes in h-BN defect states will further transfer to $ReS_2$ since a type-I band alignment is formed between the h-BN defect states and $ReS_2$. The photogenerated current is tunable under a global gate bias as shown in **Figure 3c** under different drain biases. It should be noted that the photocurrent slowly increases as the gate bias increases from -60 V to +60V (**Figure 3d**).

Next, we investigate logic gate operation using parallel MIS-type diodes with a low turn-on voltage, which can be used for more complex and functional electronic devices. It should be mentioned that an assembly of more MIS-type diodes would produce a multiple-input-based OR gate logic operation.

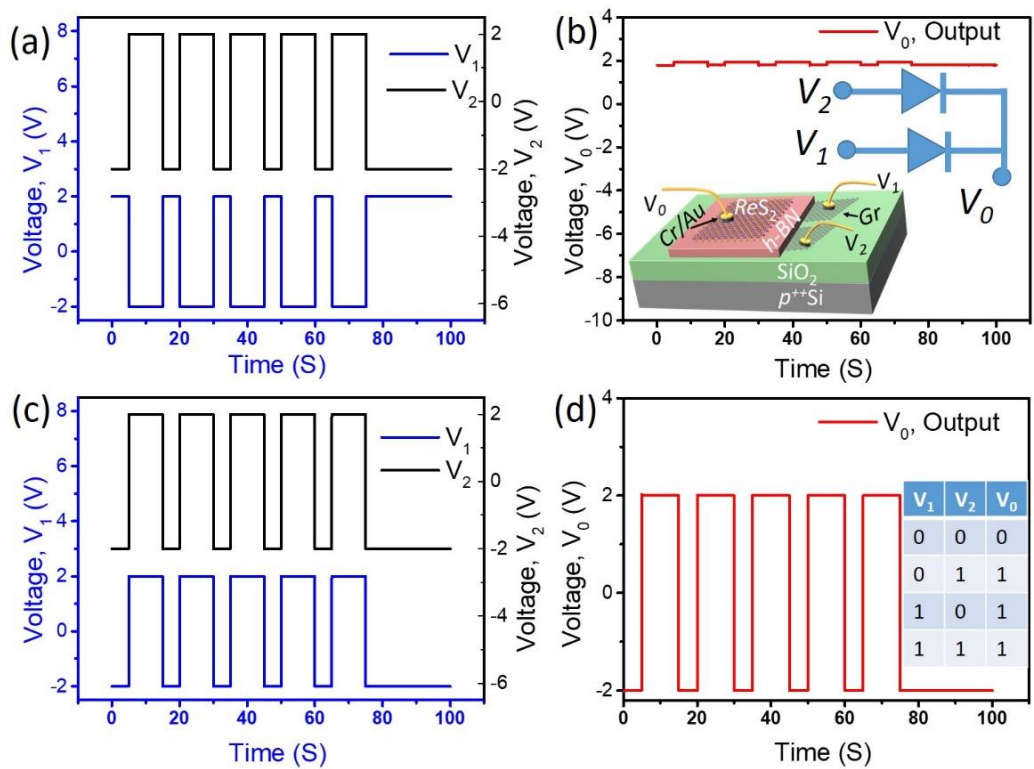

**Figure 4.** Logic gate operation using two MIS-type diodes: (a, b) Time domain plot of two different inputs at two different graphene drain terminals and the output voltage at the common source terminal of ReS₂ for OR logic gates, respectively. The inset is a schematic of the electrical connections of the dual input logic OR circuit to the device. (c, d) Time domain plot of two identical inputs and output, respectively. The device demonstrates logic OR functionality, which is controlled by independently addressing each drain terminal at graphene and the output voltage from the common source at ReS₂ terminal. The insets show a three-dimensional schematic of a multilayer ReS₂/h-BN/graphene MIS-type diode and the truth table of the OR logic operation of the device, where $V_1$ and $V_2$ are the (gate) inputs and $V_0$ is the output.

We demonstrated a simple circuit, where tunnelling diodes are connected in parallel to the output voltage monitor to form logic circuits. 'OR' logic gate operation is demonstrated. The input bias

Article type: Full Paper

at the diodes can be either high or low and the obtained output is always high (**Figure 4a,b**). Three-dimensional (3D) schematic (**Figure 4b** inset) for the diode with a $V_{in}$ drain terminal at graphene, where a pair of ReS$_2$/h-BN/graphene diodes with a common source terminal at the ReS$_2$ is at the $V_o$ monitor along with another diode in parallel as shown in the inset of the **Figure 4b** schematic measurement circuit. The output voltage levels were monitored with the four possible logic address level inputs: (0,0); (0,1); (1,0); (1,1). Here, the logic 0 input is -2 V and the logic 1 input is +2 V. In this circuit, the output is low (logic 0) when both input voltages are low (-2 V), and the output is high (logic 1), when either or both of the input voltages are high (+2 V). The output voltage levels (**Figure 4b,d**) resemble the 'OR' logic gate operation as shown **in the Figure 4d inset** truth table. As different diodes have different ON currents, turn-ON voltages and/or ON (forward)/OFF (reverse) current ratios, the `1` state of the logic output voltage shows a slight periodic modulation with time (**Figure 4b**). Small modulations in the output voltage at a high level (logic 1, **Figure 4b**) do not affect the basic operation of the logic gates because the low turn-on voltage contributions are reproducible and can be readily accounted for when defining the 0 and 1 logic states. If both inputs are logic low or high, which change periodically as shown in **Figure 4c**, then the output voltage (**Figure 4d**) we record is low or high in the same frequency and it alters in accordance with the input frequency. We have shown the 'OR' logic gate operation for multiple inputs in S.I. (**Figure S4**). Similarly, another logic gate ('AND' logic) operation can be demonstrated as shown in **Figure S5**. Due to high rectifying characteristics, these diodes can be used as temperature independent logic gate components.

## II. b. FETs at Low Threshold Voltage & Memory Operation Demonstration:

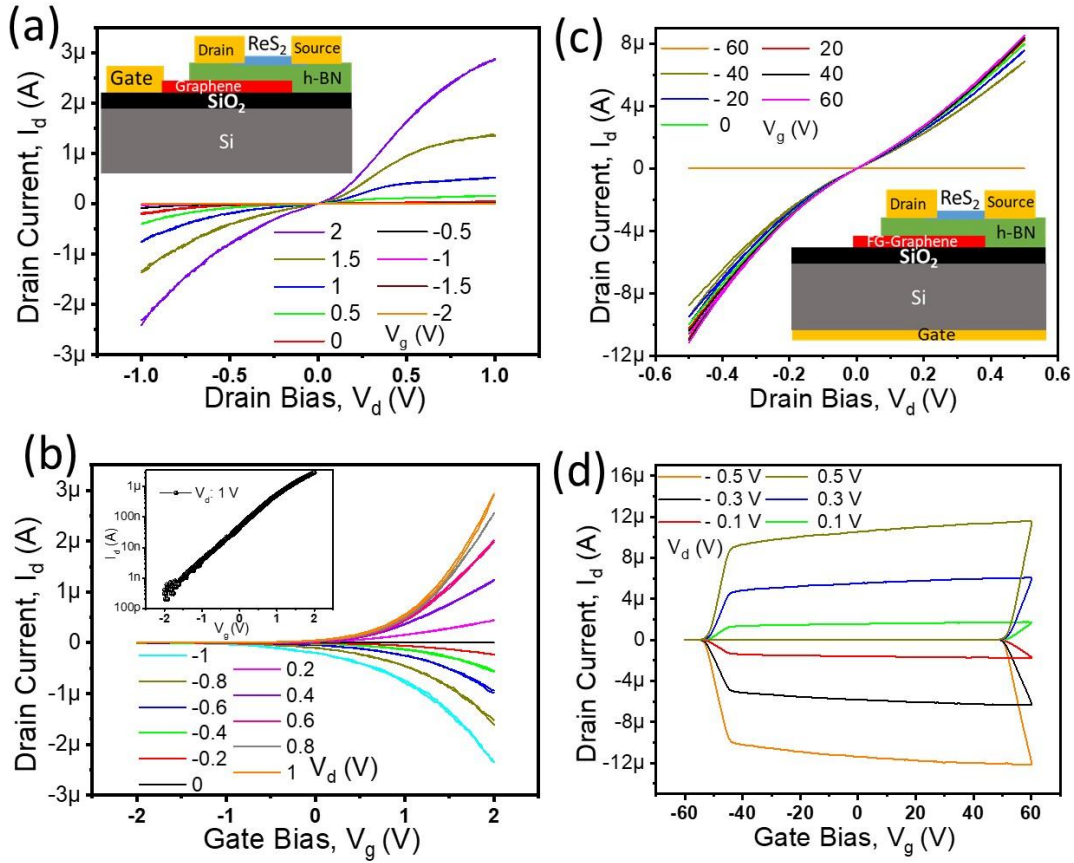

**Figure 5.** Low operating bias FET operation: electrical performance of the ReS$_2$ FET with local graphene as the bottom gate electrode as shown in the Figure 5a inset: (a, b) $I_d$-$V_d$ output characteristics and $I_d$-$V_g$ transfer characteristics of the transistor, respectively. $I_d$-$V_g$ at fixed $V_d$ of 1V is shown on a logarithmic scale plot in Figure 5b inset. NVM device operation: (c, d) Sweep $I_d$-$V_d$ output characteristics and sweep $I_d$-$V_g$ transfer characteristics of the graphene floating gated transistor as shown in the schematic diagram in Figure 5c.

A similar vertical heterostructure can be further explored for various electronic devices including a FET with low bias operation and an NVM. Here, we further explored the possibility of using the same device structure by changing and arranging the electrode connection in the circuit. Various applications including diodes, logic gates, transistors, and memory operation confirm the possibility of using a multifunctional device, whose individual functionality can be tuned further under laser light excitation to allow the possibility of use in the optoelectronics domain. First, a few-layer h-BN gate dielectric with a thickness of ~9 nm is used for ReS$_2$ FET operation, where the bottom graphene electrode is used as a gate terminal. The local gated ReS$_2$ FET performance, i.e. the transfer and output characteristics, are shown in **Figure 5 (a,b)**. As the positive gate bias at graphene terminal is gradually increased, the Fermi level enters the higher conduction band of a large density of states indicating the strong capacitive coupling of the chemical potential of graphene to the electronic band in the few layer ReS$_2$ channel (**Figure 5a**). This ReS$_2$ transistor exhibits a clear n-type characteristic with an ON/OFF current ratio of ~$10^4$-$10^5$. The output characteristics are slightly non-linear in the low bias region, which could be due to the formation of asymmetric contact and their asymmetric gate field modulation. Notably, the gate modulated drain current was observed at a low gate bias of 2V, which makes it possible to demonstrate ReS$_2$ FETs that operate at low power. A very low negative threshold voltage (-0.16 V) further confirms the natural n-doping conductance of the few layer ReS$_2$ channel, which is advantageous in terms of building a CMOS circuit for low power operation. ReS$_2$ FETs with h-BN as their back gate dielectric provide superior transistors with negligible hysteresis in the transport characteristics. This could be due to the atomically flat 2D surface of the dielectric and channel material with a 2D atomically thin gate with the absence of interface defect states. Compared with the $p^{++}$Si back-gated few-layer ReS$_2$ channel-based FET device (**Figure 2a**), this graphene/h-BN/ReS$_2$ FET device has a high threshold voltage for low power operation.

On other hand, we can use same heterostructure as an NVM device as shown in **Figure 5 (c,d)**. In this memory application, graphene acts as a floating gate that stores charges from the ReS$_2$ channel material. With global gate $p^{++}$Si control FET operation, the sweep-gate-bias-dependent drain current exhibits a large hysteresis (**Figure 5d, Figure S6** on logarithmic scale), which indicates a memory operation with a large memory window that depends on the applied gate bias value. Here a high positive and negative gate electric field with a small drain bias induced carrier tunnelling across a thin h-BN layer (~9 nm thickness) to store and erase carriers in the floating gate graphene layer and thus demonstrate the two memory states of the device operation, namely write/program (binary '1') and erase (binary '0'). Good data retention properties of the NVM devices were achieved in the time range of ~ $10^4$ s (**Figure S7** in S.I.). The transfer characteristics ($I_d$-$V_g$) for different gate bias ($V_{g,max}$) ranges represent large hysteresis windows, which can be tuned by the maximum gate bias (**Figure S8** in S.I.). The details of NVM operation with extended multilevel operation under laser excitation with a similar device structure have been reported elsewhere.[25] Here, we achieved NVM operation under a moderate electric field with a +/-10V gate bias, which can be further reduced if we use high quality h-BN as the dielectric layer for the control gate ($p^{++}$Si) instead of using a $SiO_2$ dielectric layer. For an all 2D-material-based NVM device with low gate bias operation, the FG-graphene can be isolated by both a top h-BN (~ 6-10 nm thick) and a bottom h-BN (~ 30-40 nm thick) subsequently insulated with another bottom graphene as a control gate electrode. That multi-stack heterostructure will help us to realize a NVM memory device operating at very low power for multibit applications, which will be further studied in future work with extra functionality and a laser tunable optoelectronic NVM device. We have discussed various uses starting with vertical MIS-type diodes, low power transistor operation, logic gate operation and memory device applications for future ultrathin nanoelectronics based multifunction devices.

## III. Conclusion

We described various multifunctional electronic devices with an $ReS_2$ based 2D heterostructure. We reported a vertical metal-insulator-semiconductor (MIS)-like configuration as a quantum tunnelling diode structure and the logic gate operation when using tunnel diodes with a detailed characterization of the gate electric field dependence, temperature dependence and incident light excitation. The vertical geometry can lead to the high rectification of tunnel devices with an atomically flat 2D surface as a high tunnelling area for a large turn-ON current. The fabricated parallel diodes facilitate both direct tunnelling (DT) at a lower field strength and Fowler-Nordheim (FN) tunnelling with the use of a high-strength electric field, which we used to design logic gates that depended on the various combinations we used. We obtained a large rectification, low ideality factor, temperature independent tunnel diodes and their `OR` logic gate operation. We demonstrated $ReS_2$ FETs operating at a low bias and an NVM device with a large memory window in the same heterostructure device. These devices are candidates for future 2D ultrathin electronic devices, namely diodes, transistors and memory devices, as building blocks for logic circuits in next generation VLSI architectures, which can potentially enable all-optical logic processing and quantum information processing in the modern smart Internet-of-Things (IoT) technology.

## IV. METHODS

**Device Fabrication and Measurement Techniques:**

Standard dry-transfer techniques[31,32] were used to prepare a heterostructure consisting of $ReS_2$, h-BN and graphene. Initially a $SiO_2$ (285 nm)/Si substrate was used, which was treated with oxygen plasma to remove contaminants and to induce surface modification with functional silanol groups (Si-OH).[28] Then mechanically exfoliated graphene flakes from highly oriented pyrolytic graphite

Article type: Full Paper

(HOPG) were deposited on the surface of the plasma treated $SiO_2$ at a substrate temperature of 180 $^0$C. Next an h-BN layer and $ReS_2$ flakes were transferred to the top of the graphene/$SiO_2$ sample using the dry transfer procedure. Standard electron beam lithography (EBL, ELS-7000, F125 KV) techniques were used to pattern the electrodes on the fabricated heterostructure device. An electron beam sputtering unit was used to deposit Cr/Au (Cr: 5 nm in 0.03 nm/s deposition rate, Au: 50 nm, 0.15 nm/s deposition rate in a high vacuum of $10^{-5}$ Pa).

All electrical measurements were performed with the standard two-probe and three-probe measurements configuration. The current–voltage ($I$–$V$), current–time ($I$–$t$) and all the optoelectrical characteristics (photocurrent-time) of the device were measured using an Agilent 2636A and a semiconductor device analyser (Agilent B1500A) source-measurement unit. The devices were tested in a high-vacuum chamber ($5 \times 10^{-3}$ Pa) in a Lakeshore probe station at room temperature. A diode laser (532 nm, diode-pumped solid-state DPSS laser) was used for the optoelectrical response. An atomic force microscope (Nikon/SHIMADZU, model SPM-9700HT, scanning probe microscope) and a Raman microscope (Nanophoton, model Ramanplus, 532 nm laser, with ×100 - 0.9 N.A. objective lens and 1200 lines/mm grating) were used for the thickness measurement and sample characterisation, respectively.

**ACKNOWLEDGMENT**

This research was supported by the World Premier International Center (WPI) for Materials Nanoarchitectonics (MANA) of the National Institute for Materials Science (NIMS), Tsukuba, Japan with a Grant-in-Aid for Scientific Research (JSPS KAKENHI Grant No./Project/Area No.17F17360). A part of this study was supported by the NIMS Nanofabrication Platform and

**Competing interests**

The authors declare no competing interests.

**Additional information**

**Correspondence** should be addressed to B.M. or Y.W.

**ORCID**

**Bablu Mukherjee:** 0000-0002-5625-5948

**Ryoma Hayakawa:** 0000-0002-1442-8230

**Kenji Watanabe:** 0000-0003-3701-8119

**Takashi Taniguchi:** 0000-0002-1467-3105

**Shu Nakaharai:** 0000-0002-6329-3942

**Yutaka Wakayama:** 0000-0002-0801-8884

Article type: Full Paper

ToC Figure:

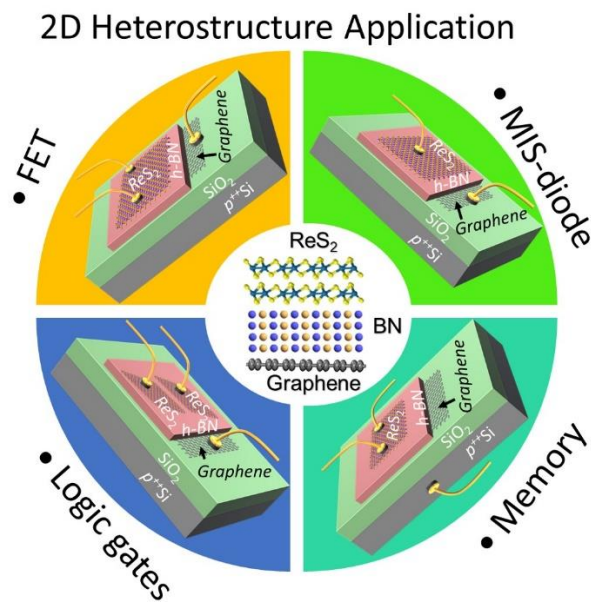

Significance:

Two-dimensional van-der-Waals-heterostructure-based multi-functional device. Various applications including MIS-type diodes, field effect transistors, logic gates and non-volatile memory devices are demonstrated. A low ideality factor with high rectification are achieved in diodes, while a low threshold voltage is demonstrated in transistor operation. These active components can be used in simple logic gate circuits and in high memory window devices.

# *Supporting Information*

# ReS₂/h-BN/Graphene Heterostructure Based Multifunction Devices: Tunnelling Diodes, FETs, Logic Gates & Memory

Bablu Mukherjee[1*], Ryoma Hayakawa[1], Kenji Watanabe[2], Takashi Taniguchi[1], Shu Nakaharai[1], Yutaka Wakayama[1*]

[1]International Center for Materials Nanoarchitectonics (MANA), National Institute for Materials Science (NIMS), 1-1 Namiki, Tsukuba, Ibaraki 305-0044, Japan.

[2]Research Center for Functional Materials, National Institute for Materials Science (NIMS), 1-1 Namiki, Tsukuba, Ibaraki 305-0044, Japan.

* Corresponding authors. Email: MUKHERJEE.Bablu@nims.go.jp (Bablu Mukherjee); WAKAYAMA.Yutaka@nims.go.jp (Yutaka Wakayama)

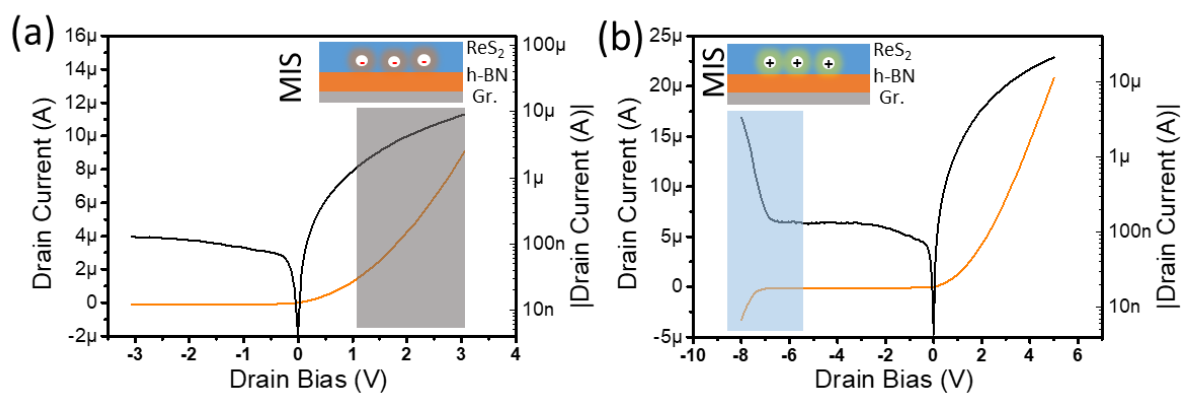

**Figure S1:** (a,b) Room temperature electrical output characteristics of an MIS-type diode for different bias ranges. Here, the black lines are plotted on linear scale and the red lines are shown on logarithmic scale, respectively. Electron accumulation occurs at the interface ReS₂ and h-BN layers when the positive bias voltage is applied to the graphene electrode, as shown by the grey shaded region in (a) and schematically shown in the inset (a). Weak hole inversion occurs at the negative bias voltage as shown by the inset schematically and by the shaded blue region in figure (b).

Article type: Full Paper

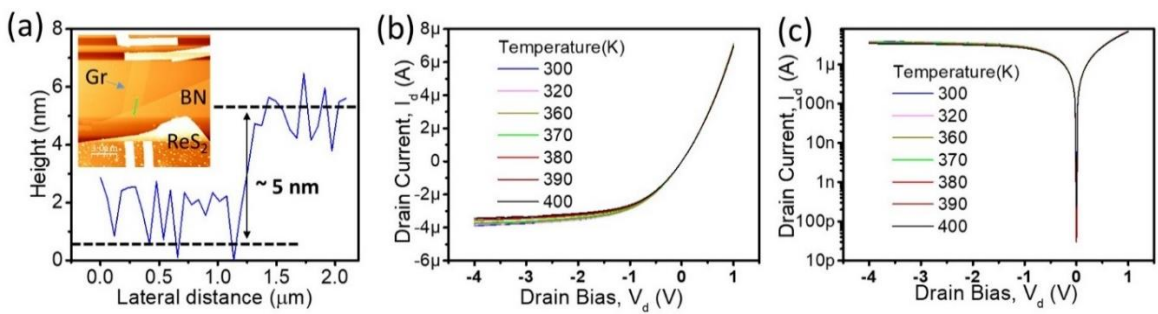

**Figure S2**: (a) Line profile of the thickness across the green line of the h-BN flakes. Inset shows surface topography AFM image of MIS-type diode device. Temperature dependent electrical characteristics of MIS-type tunneling diode: (b, c) Current *versus* voltage with different temperatures of the MIS-type two-probe device structure on linear and log scales, respectively. The $ReS_2$ terminal was fixed to a drain and the graphene terminal was connected to a source.

| Intensity (mW/cm$^2$) | $I_{Forward}$ (µA) | $I_{Reverse}$ (µA) | Ratio ($I_{Forward}/I_{Reverse}$) |
|---|---|---|---|
| 0 | 3.6 | 0.7 | 5.1 |
| 3 | 5.9 | 0.72 | 8.2 |
| 7 | 6.65 | 0.77 | 8.6 |
| 10 | 9.55 | 0.8 | 11.9 |

**Table 1:** Rectification ratio of the MIS-type diode with different light intensities.

Article type: Full Paper

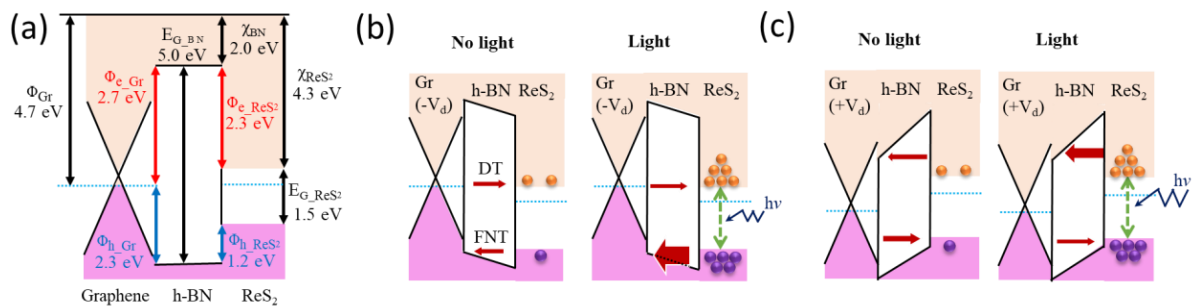

**Figure S3:** (a) Flat energy band diagram of vertically integrated ReS$_2$/h-BN/graphene memories, where χ, φ, and E$_g$ represent the electron affinity, work function, and band gap, respectively. (b) Energy band diagram of ReS$_2$/h-BN/graphene at negative drain voltage at graphene terminal with and without light illumination and (c) Energy band diagram of ReS$_2$/h-BN/graphene at positive drain voltage at graphene terminal with and without light illumination.

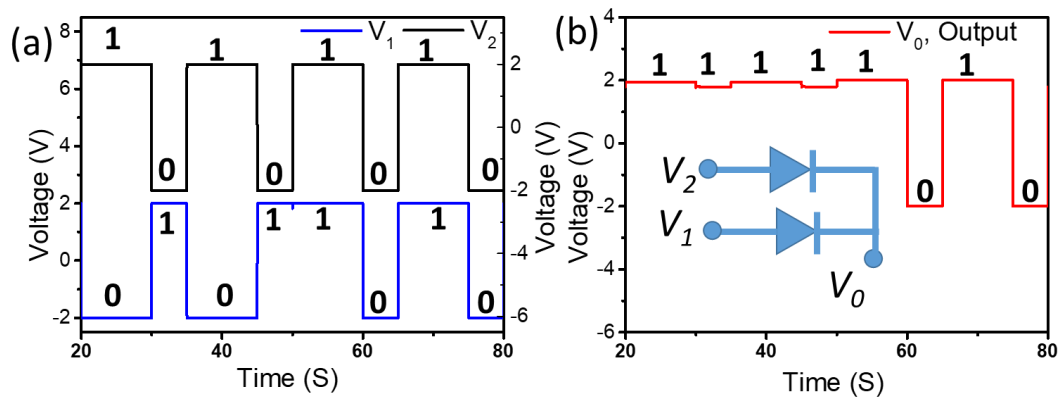

**Figure S4**. Logic OR gate operation using two MIS-type diodes: (a, b) Time domain plot of two multiple inputs at two different graphene drain terminals and the output voltage at the common source terminal of ReS$_2$ for OR logic gates, respectively. The inset is a schematic of the electrical connections of the dual input logic OR circuit to the device.

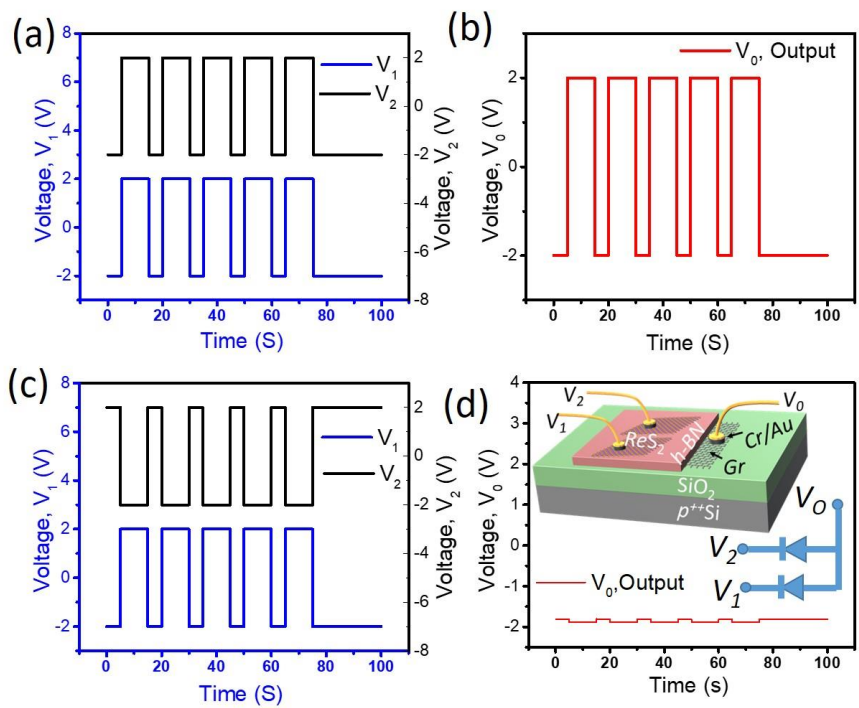

**Figure S5**: AND logic gate operation using two MIS-type diodes. (a) Pulsed inputs with the same phase, amplitude and frequency at the diodes. (b) The output at the common terminal of graphene, which is the same as the inputs. (c,d) Inputs and output from the logic circuit with opposite input pulses. The insets show a schematic of the AND logic circuit and the device.

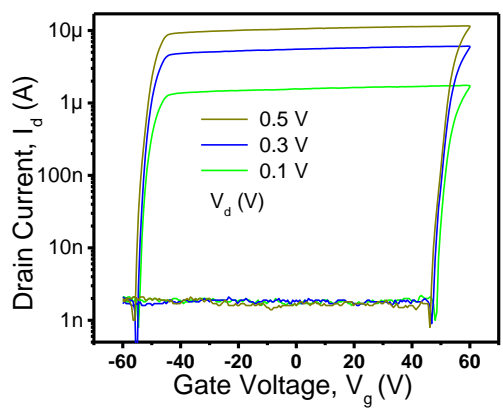

**Figure S6**: $I_d$-$V_g$ transfer characteristics of the graphene floating gated $ReS_2$ transistor on a logarithmic scale.

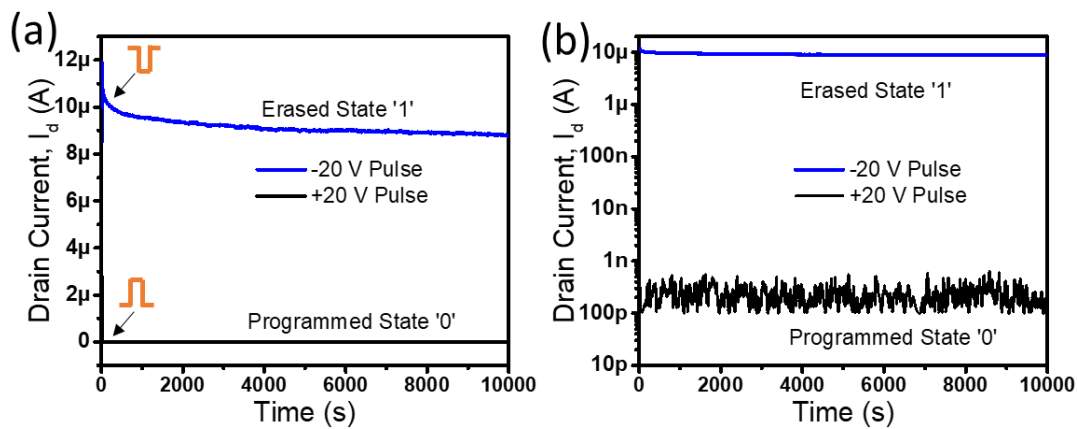

**Figure S7**. Retention time characteristic of $I_d$ in the ON and OFF states on a (a) linear scale and (b) logarithmic scale, respectively. Each state was read at $V_g = 0$ V, $V_d = 0.5$ V after being programmed (erased) by one pulse voltage of $+ 20$ V ($-20$ V) and a width of 5 s on the control gate.

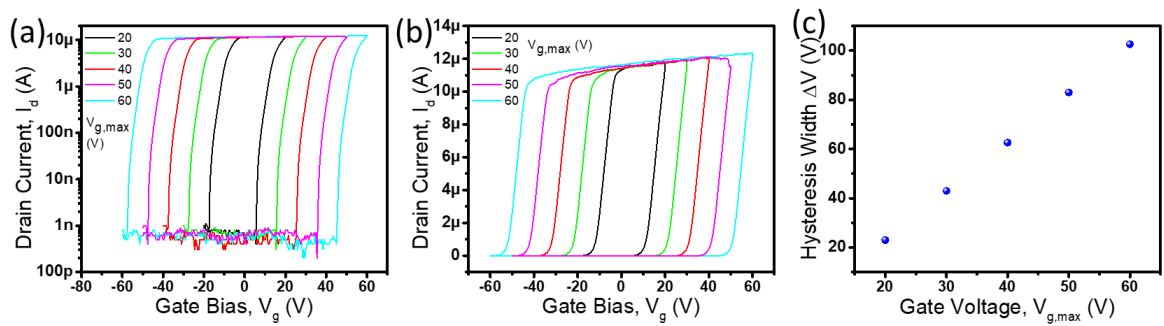

**Figure S8**: (a) $I_d$-$V_g$ curves with different $V_{g,maximum}$ represent large hysteresis width on linear scale and logarithmic scale, respectively. (c) Hysteresis width *versus* $V_{g,max}$.

Article type: Full Paper

**Threshold voltage calculation:**

The threshold voltage of the device was calculated from linear region in $I_d$-$V_g$ curves by the following equation:

$$I_d = \frac{W}{L}\mu C_i \left[ (V_G - V_{th})V_d - \frac{1}{2}V_d{}^2 \right]$$

Where $L$=1.2 μm and $W$=8.4 μm are the length and width of the device channel, respectively. $C_i$=$\varepsilon_0\varepsilon_r/d$ is capacitance per unit area calculated for the gate dielectric (h-BN), with $\varepsilon_0$=8.85 x10$^{-12}$ F/m is the free-space permittivity, and $\varepsilon_r \sim 3.9$ is the relative permittivity of h-BN. Meanwhile, $d$=10 nm is the h-BN thickness. The calculated $C_i$ for 10 nm h-BN is $\approx 345$ nF/cm$^2$. μ, $V_{GS}$, $V_{th}$, and $V_{DS}$ are mobility, gate voltage, threshold voltage and drain voltage, respectively.